\documentclass[aip,rsi,reprint,superscriptaddress,letterpaper]{revtex4-2}

\usepackage{amsmath}
\usepackage{amssymb}
\usepackage{braket}
\usepackage{mathtools}
\usepackage{bm}
\usepackage{verbatim}

\begin{document}

\title{3D-Printed Enclosure Wire-Guided Liquid Microfilm for Versatile Spectroscopy}
\author{Matthew J. Silverstein}
\email{mattsil@umich.edu}
\affiliation{Department of Chemistry, University of Michigan, Ann Arbor, MI 48109, USA}
\author{Yasashri Ranathunga}
\email{yasashri@umich.edu}
\affiliation{Department of Chemistry, University of Michigan, Ann Arbor, MI 48109, USA}
\author{Yuki Kobayashi}
\email{ykb@umich.edu}
\affiliation{Department of Chemistry, University of Michigan, Ann Arbor, MI 48109, USA}

\date{\today}

\begin{abstract}
We present a 3D-printing-based design to produce wire-guided liquid microfilms that can be used for versatile spectroscopic applications.
We demonstrate the ability of our instrument to provide optically useful liquid microfilms with highly tunable thicknesses over the range 25 -- 180 \textmu m, with standard temporal thickness deviation less than 1.0\% on the low end of the range of flow rates, and spatially homogeneous microfilms that remain stable over the course of ten hours.
We then show the device’s versatility through its use in Raman, fluorescence, and nonlinear spectroscopy.
Our approach is highly reproducible as a unique advantage of a 3D-printed enclosure and limited other components.
The 3D-printable file for the enclosure is included in the supplementary materials.
This innovation in design shows the feasibility of applying 3D-printing to physical and chemical instrumentation for faster adoption of experimental techniques. 
\end{abstract}

\maketitle
%%%%%%%%%%%%%%%%%%%%%%%%%%%%%%%%%%%%%%%%%%%%%%%%%%%%%%%%%%%%%%%%%%%%%%%%%%%%%%%
%%%%%%%%%%%%%%%%%%%%%%%%%%%%%%%%%%%%%%%%%%%%%%%%%%%%%%%%%%%%%%%%%%%%%%%%%%%%%%%
\section{Introduction}
Wire-guided liquid microfilms offer a desirable answer to many of the problems faced in solution-phase spectroscopy.
By eliminating media external to the sample and reducing the sample pathlength to a minimum, many obstacles related to dispersion, absorption, and optical nonlinearities can be mitigated.
For example, terahertz radiation is absorbed by materials such as glass,\cite{Naftaly2007} but this loss is mitigated by using the window-less liquid films.
Ultrashort laser pulses can be temporally stretched by dispersion and this is solved by the elimination of external media in the optical path and reduction of sample pathlength.
Lastly, high-intensity lasers can induce sample or sample holder damage and undesired optical nonlinearities such as the optical Kerr effect, which can be solved by using window-free liquid films.

The first wire-guided liquid microfilm apparatus was introduced in 2003\cite{Tauber2003} for use in resonance Raman and ultrafast spectroscopy, achieving very low sample path lengths (6–100 \textmu m) with minimal deviations in thickness.
The design used a peristaltic pump and stainless steel tubing to create a liquid microfilm supported with surface tension by thin metal wires.
Later, another design was introduced\cite{Picchiotti2015} that altered the previous design by replacing the peristaltic pump with a microfluidic pump and adding a damper to minimize low-frequency fluctuations.
Both of these designs, as well as other designs to produce liquid films and jets\cite{Ekimova2015,Riley2019,Yiwen2018} offer great successes, and several laser spectroscopy groups have developed setups utilizing similar wire-guided microfilm instruments in their optical experiments.\cite{Wang2014,Jin2017,Kuramochi2016,Huang2024}

The need for liquid microfilms is more than ever before as ultrafast spectroscopy is becoming more common,\cite{Loh2020,Low2022,Yin2023} terahertz spectroscopy is gaining popularity,\cite{Wang2014,Zhao2020} and liquid microjets are making solution-phase photoelectron spectroscopy possible.\cite{Stemer2023,Kang2024}
New optical uses of liquid microfilms and jets are being found, such as for supercontinuum generation,\cite{Huang2024} terahertz generation,\cite{Yiwen2018,Jin2017} deep-ultraviolet pulse characterization,\cite{Rivera2010} and even high-harmonic generation.\cite{Luu2018,Alexander2023}
In addition to their usage in physical chemistry, liquid microfilms have found applications in analytical chemistry,\cite{Lo2021} electrochemistry,\cite{Srivastava2024} biophysics,\cite{Mizutani2017,Perlk2015,Donten2013} and the study of nanomaterials.\cite{Seferyan2006,Sperling2010}
Liquid microjets are even finding use outside of optics and spectroscopy in novel experiments on molecular beam scattering.\cite{Lee2022,Yang2024}
However, despite their well-regarded utility, wire-guided liquid microfilm instruments are difficult to reproduce without the expertise of machine shops, and the community would benefit from a design that can be easily copied and installed.

3D-printing using plastics has become valuable for making custom parts for spectroscopy.\cite{Dewberry2015,Grasse2016,Wilkes2017,Baumgartner2020,daSilvaJunior2021}
For a sample holder directly in contact with chemicals, however, plastics typically used for 3D-printing such as polylactic acid can degrade.\cite{Elsawy2017}
Metal 3D-printing has been applied to electrode fabrication for catalysis and electrochemistry,\cite{Ambrosi2016,Wei2020} but has remained unexplored as a technology for the development of spectroscopic instrumentation.

Here, we show a metal 3D-printed design to produce wire-guided liquid microfilms in order to address the aforementioned problems of solution-phase spectroscopy and ease of reproducibility.
Our device uses a 3D-printed enclosure with no custom-machined parts so that interested researchers can easily employ wire-guided liquid microfilms in their optical setups.
The whole apparatus is small, fitting within an 8" by 8" area, and the enclosure is mounted on an optical post for versatility in either low-space stationary positions, or on translation stages.
Throughout the following sections, we detail the design, characterization, and demonstrated applications of our high-accessibility wire-guided liquid microfilm instrument.

%%%%%%%%%%%%%%%%%%%%%%%%%%%%%%%%%%%%%%%%%%%%%%%%%%%%%%%%%%%%%%%%%%%%%%%%%%%%%%%
%%%%%%%%%%%%%%%%%%%%%%%%%%%%%%%%%%%%%%%%%%%%%%%%%%%%%%%%%%%%%%%%%%%%%%%%%%%%%%%
\section{Apparatus}
\begin{figure}[tb]
\includegraphics[scale=1]{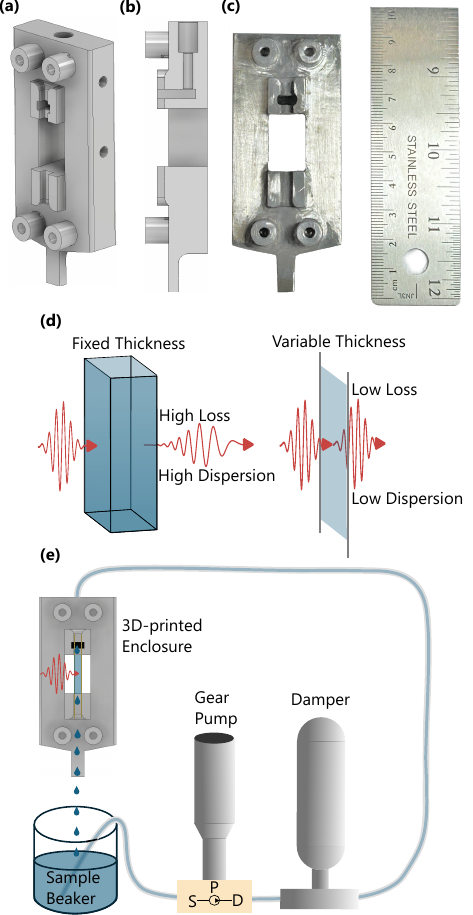}
\caption{\label{fig1}
{\bf Schematics of the wire-guided liquid microfilm apparatus}
{\bf (a)} Digitally rendered view of the 3D-printed enclosure.
{\bf (b)} Cross-sectional view of the enclosure model, showcasing the liquid-flow pathway approaching the wire-guide.
{\bf (c)} Image of the actual 3D-printed part, with a ruler for scale.
{\bf (d)} Illustration of the benefits of a wire-guided liquid microfilm over an ordinary cuvette.
{\bf (e)} Overview of the whole apparatus with all functional components.
}
\end{figure}

Our wire-guided liquid microfilm setup consists of a main 3D-printed body, thin metal wires, tube fittings, a damper, and a microannular gear pump.
The CAD file for the 3D-printed body is included in the supplementary materials.
A schematic of the system is shown in figure \ref{fig1}(e).

The 3D-printed enclosure has many quality-of-life features that aid in its utility as a sample holder for chemical spectroscopy.
First, 3D-printing allows for fast and accurate emulation of the design by any interested researchers, without the need for access to a machine shop or custom parts.
Another direct consequence of the design being 3D-printable is the ability to swiftly replace a damaged or otherwise ineffective enclosure with an identical copy.
Additionally, 3D-printing allows for the device to be made with a metal that fits the needs of a particular chemical sample.
The form factor of the main body makes it easily removable from the rest of the setup in order to clean off chemical residues or exchange the thin metal wires.

Flowing liquid sample holders are known to exhibit temporal thickness deviation from the pump mechanism.\cite{Kuramochi2016}
To combat this, the top of the enclosure provides an entrance to the wires that would remove most additional pump pressure or oscillatory noise prior to the liquid's gravity-driven descent.
This would be hard to incorporate using conventional machining, but is easily achieved with 3D-printing.
This feature is seen in the cross-sectional view of the 3D model in Fig. \ref{fig1}(b).

Our enclosure was designed using commercially available CAD software and was printed out of aluminum (AlSi10Mg) with layer accuracy to tens of microns.
This printing and additional post-processing to improve part quality was performed affordably by an external contractor.
The pump that we use is an mzr-2921x1 low-pressure microannular gear pump that can produce flow rates from 0.03--18 mL/min.
The pump is interfaced with the S-G05 controller; both the pump and the controller are from HNP Mikrosysteme.
The thin metal wires to guide the liquid are made from 0.003" diameter tungsten; this is comparable to the wires used in the first reported wire-guided liquid microfilm.\cite{Tauber2003}
We also make use of a damper fashioned from a Swagelok sample cylinder with a volume of 40 cm$^3$ to further reduce fluctuations in fluid flow.
Connections between all the components are made by silicone tubing with an inner diameter of 2 mm and an outer diameter of 4 mm.

%%%%%%%%%%%%%%%%%%%%%%%%%%%%%%%%%%%%%%%%%%%%%%%%%%%%%%%%%%%%%%%%%%%%%%%%%%%%%%%
%%%%%%%%%%%%%%%%%%%%%%%%%%%%%%%%%%%%%%%%%%%%%%%%%%%%%%%%%%%%%%%%%%%%%%%%%%%%%%%
\section{Characterization}

When working with a chemical sample holder for use in optical spectroscopy, not only is it important that a low pathlength is achieved, but also that the thickness is precisely known and that the deviation in the pathlength is minimal.
In this section, we provide optical characterization of the film performance in our design.
The approach is based on UV-visible absorption by potassium ferricyanide dissolved in water.
Using a 2-mm-pathlength cuvette and a photodiode, we determined that the molar absorption coefficient at 405 nm is 990 M$^{-1}$cm$^{-1}$, and this value is used in the following analysis (see supplementary materials).

\begin{figure*}[tb]
\includegraphics[scale=1]{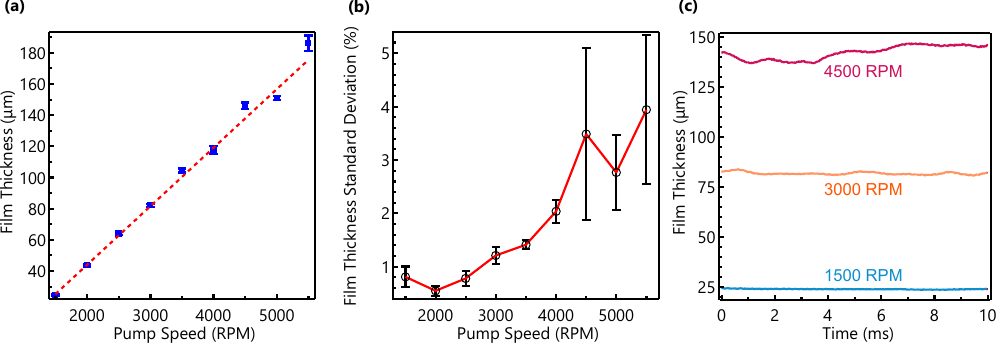}
\caption{\label{fig2}
{\bf Static position measurements of the wire-guided liquid microfilm thickness and stability at various gear pump speeds}
{\bf (a)} Liquid microfilm thickness as a function of gear pump speed, as measured by the average between nine repetitions (three at each of three concentrations of chromophore) of $1\times10^5$ photodiode samples with a sampling rate of 100 kHz. Error bars are the standard deviation between repetitions at the same RPM.
{\bf (b)} Percentage standard temporal deviation of the wire-guided liquid microfilm, as determined by the average of all of the standard deviations of the one-second measurements for a given RPM. Error bars are the standard deviation between repetitions at the same RPM.
{\bf (c)} Thickness of the microfilm over a 10 ms segment for pump speeds of 1500, 3000, and 4500 RPM.
}
\end{figure*}

\subsection{Static Position Measurements}
We first characterized the thickness of the produced liquid microfilm as a function of the pump speed.
By focusing a continuous-wave 405 nm laser on the liquid microfilm, and recording the photodiode signal with and without potassium ferricyanide in the water flowing through the device, we were able to measure the thickness according to the Beer-Lambert law and our recorded molar absorption coefficient of potassium ferricyanide.
Values for the absorbance were collected by taking the average of nine repetitions (three at each of three concentrations of chromophore: 30, 40, and 50 mM) of $1\times10^5$ samples taken at a sampling rate of a data acquisition system at 100 kHz.

As can be seen in Fig. \ref{fig2}(a), a linear relationship can be drawn between the pump speed and the liquid microfilm thickness over the range from 25 -- 180 \textmu m corresponding to pump speeds of 1500 RPM and 5500 RPM, respectively.
This shows that we have a predictably tunable liquid microfilm thickness and that we could tune the thickness to a desired value simply by changing the pump speed.

The optical measurements also allow for the extraction of a temporal deviation in the thickness of the wire-guided liquid microfilm.
Standard deviations were calculated for each one-second measurement and averaged for all of the measurements at a given RPM.
As can be seen in Fig. \ref{fig2}(b), the liquid microfilm achieves small temporal deviations of thickness, which stay below even 1.0\% on the low end of the range of gear pump speeds.
Thickness fluctuations become noticeable above 2500 RPM, and the deviation between replicate measurements (as denoted by error bars) becomes quite large, spanning greater than $\pm$0.5\% above 4000 RPM.

It is also important to note that over multiple trials, the microfilm ran with a pump speed of 2500 RPM for ten hours with no observed breakage using photodiode measurement.
Below this pump speed there is occasional microfilm breakage.

Overall, it is clear from static position measurements that our wire-guided liquid microfilm instrument meets the standards of an optically useful sample holder for various modes of spectroscopy, with high tunability for greater sample control.

\subsection{Spatial Deviation}
\begin{figure*}[tb]
\includegraphics[scale=1]{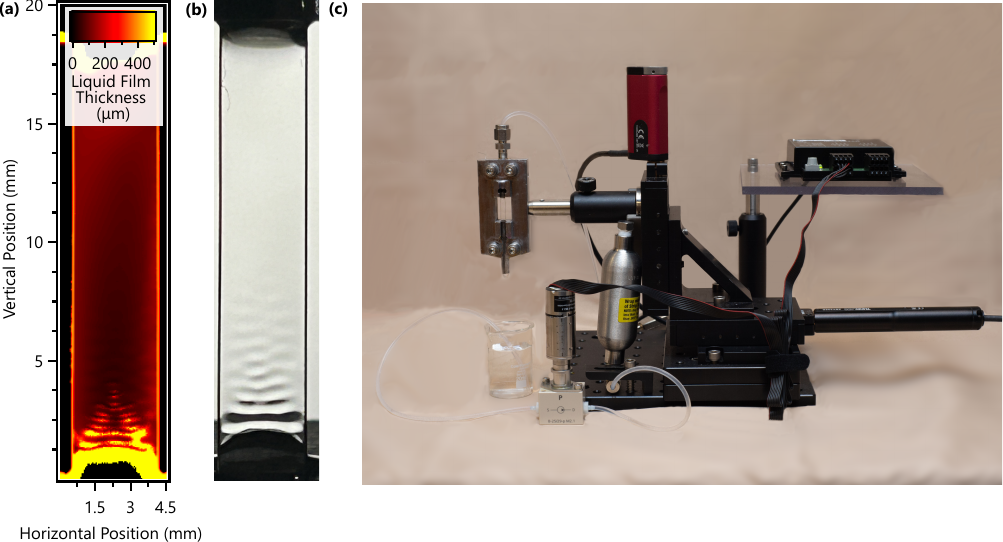}
\caption{\label{fig3}
{\bf Spatial profiling of the wire-guided liquid microfilm}
{\bf (a)} 2D thickness map of the wire-guided liquid microfilm, taken at a pump speed of 2500 RPM.
{\bf (b)} Image of the wire-guided liquid microfilm in operation.
{\bf (c)} Image of the wire-guided liquid microfilm setup with dual stepper motors used to perform 2D thickness scans.
}
\end{figure*}

We then evaluated the spatial profile of the wire-guided liquid microfilm by placing the 3D-printed body onto two-axis motorized stages.
For the measurement, a 405-nm laser was focused to a spot size of 130 \textmu m and the transmission was detected by a photodiode.
A vertical and horizontal step size of 50 \textmu m for stepper motor movement were used in order to capture the thickness features of the wire-guided liquid microfilm.
We again used potassium ferricyanide along with the Beer-Lambert law to compute the thickness.
The setup for this measurement is pictured in Fig. \ref{fig3}(c).

As observed in Fig \ref{fig3}(a), with a pump speed of 2500 RPM the liquid microfilm is spatially homogeneous. 
There is a large amount of optically useful area on the liquid microfilm, conservatively estimated to be 15 mm$^2$.
Analyzing 1 mm$^2$ of area near the center of the microfilm, the mean thickness was found to be 66 \textmu m with a standard deviation of 4.0\% between 50 \textmu m by 50 \textmu m pixels.
This points to the film having smaller temporal fluctuations than spatial deviations.
We also see a fringe pattern at the bottom of the microfilm that is measured to be stable during several hours of operation, a feature typically seen in laminar flows [Fig. \ref{fig3}(b)].

\section{Applications of the wire-guided liquid microfilm}
Having characterized the spatial and temporal profile of the produced wire-guided liquid microfilms, we turned to demonstrating its applicability to various spectroscopic methods: Raman spectroscopy, photoluminescence spectroscopy, and two-photon absorption spectroscopy.

\begin{figure*}[!htb]
\includegraphics[scale=1]{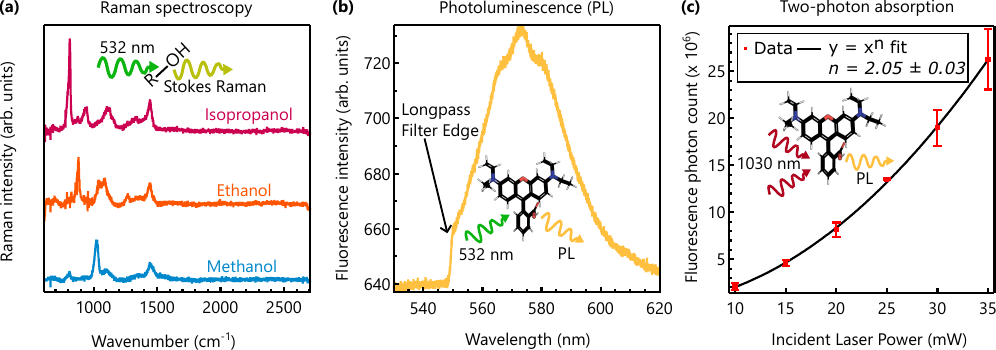}
\caption{\label{fig4}
{\bf Exemplary applications of the wire-guided liquid microfilm to various modes of spectroscopy}
{\bf (a)} Raman spectra of isopropanol, ethanol, and methanol using a 532-nm pump.
{\bf (b)} Fluorescence spectrum of Rhodamine B with a 532-nm excitation.
{\bf (c)} Nonlinear optical response of Rhodamine B to femtosecond 1030-nm excitation, as measured by the fluorescence photon count compared to the infrared input power.
A very near-quadratic nonlinearity is suggested by the fit parameter to these points, showing the expected result for a two-photon absorption process.
}
\end{figure*}

Using a 33 mW output from a continuous-wave 532 nm laser and a 5-minute integration time on the spectrometer, we were able to obtain Raman spectra from three alcohol molecules.
We freely flowed the alcohols through the apparatus at 6000 RPM.
The spectra of the alcohols after continuous background subtraction and outlier elimination are shown in Fig. \ref{fig4}(a).
Thus, the wire-guided liquid microfilm instrument is able to be used for nonresonant continuous-wave Raman spectroscopy. Furthermore, this work shows that the apparatus is capable of experiments using organic solvents.

We took a static fluorescence spectrum of 10 ppm aqueous Rhodamine B, a well-known organic fluorescent dye. \cite{Battula2021}
We used a wire-guided liquid microfilm to hold the sample and a 532 nm excitation wavelength at 33 mW of power.
A fluorescence spectrum, Fig. \ref{fig4}(b), was produced with only 50 ms of integration time.

A further application of our wire-guided microfilm apparatus is in nonlinear spectroscopy.
Whereas sample and sample holder damage could occur when using a glass cuvette, wire-guided microfilms constantly recycle the sample and have otherwise no media in the optical path that could be damaged. 
To exemplify the application of our device to nonlinear spectroscopy, we show two-photon photoluminescence from Rhodamine B, a result that is well characterized in the literature.\cite{Kristoffersen2014}
This is a particularly important phenomenon for organic dyes, as two-photon microscopy is becoming widely adopted.\cite{Alifu2017}
Flowing 100 ppm Rhodamine B to produce a microfilm and optically pumping it with a femtosecond pulsed 1030 nm laser (200 fs, 40 MHz, 10--35 mW), we were able to visually observe fluorescence.
Figure \ref{fig4}(c) shows the laser power dependence of the fluorescence signal.
Fitting of the results determines that the order of the response was $n = 2.05 \pm  0.03$, indicating our fluorescence indeed results from two-photon absorption of the high-intensity laser field.
The ability of our instrument to be used in a high-intensity experiment, such as the one presented here, shows just how versatile and ubiquitous our design could be in spectroscopy.

\section{Conclusion}
In conclusion, we have provided a new metal-3D-printed design for liquid microfilm instruments and demonstrated its versatility for spectroscopic applications.
The CAD file for the wire-guided liquid-microfilm enclosure is available in the supplemental materials for this publication, and the part can be made with ordinary metal 3D-printing to reproduce our setup.
The microfilms are shown to have stable and tunable thickness and we have shown that they can be used for various spectroscopic techniques.
With increased interest in ultrafast spectroscopy and sensitive terahertz and ultraviolet measurements, we envision that our 3D printed design will make a broad impact to a wide community of spectroscopists in biophysics, physical chemistry, and chemical physics.

%%%%%%%%%%%%%%%%%%%%%%%%%%%%%%%%%%%%%%%%%%%%%%%%%%%%%%%%%%%%%%%%
\section*{Acknowledgments}
This work was supported by the start-up fund from the University of Michigan.
M.J.S. was supported by the May-Walt Summer Fellowship Award through the University of Michigan.

%%%%%%%%%%%%%%%%%%%%%%%%%%%%%%%%%%%%%%%%%%%%%%%%%%%%%%%%%%%%%%%%
\section*{Author contributions}
M.J.S. performed the experiments and analysis.
Y.R. also assisted in experimentation and data analysis while overseeing the project, and Y.K. designed the apparatus and supervised and conceptualized the project. All authors contributed to the preparation of the manuscript.

%%%%%%%%%%%%%%%%%%%%%%%%%%%%%%%%%%%%%%%%%%%%%%%%%%%%%%%%%%%%%%%%
\section*{Competing interests}
The authors declare no competing interests.

%%%%%%%%%%%%%%%%%%%%%%%%%%%%%%%%%%%%%%%%%%%%%%%%%%%%%%%%%%%%%%%%
\bibliography{MS_microfilmrefs_ArXiv}

\end{document}